\documentclass[a4paper,twocolumn,twoside]{esapub}

\usepackage[numbers,square]{natbib}   %

\usepackage[]{graphicx}%

\usepackage[bookmarksnumbered=true,breaklinks=true]{hyperref}   %

\newlength{\textcol}
\newlength{\piccol}
\title{Direct Detection of Sub-stellar Companions with MIDI}   %

\author[1,4]{Peter Schuller$^\mathrm{*,}$}
\affil{Harvard-Smithsonian Center for Astrophysics, MS-20, 60 Garden Street, Cambridge, MA 02138, USA}
\author{Martin Vannier$^\mathrm{\dagger,}$}
\author{Romain Petrov}
\affil{UMR~Astrophysique, Universit\'e~de~Nice, Sophia~Antipolis, 28~Avenue~Valrose, 06108~Nice~Cedex~2, France}
\author{Bruno Lopez}
\affil{Observatoire~de~la~C\^{o}te~d'Azur, BP~4229, 06304~Nice~Cedex~4, France}
\author{Christoph Leinert}
\author{Thomas Henning}
\affil{Max-Planck-Institut~f\"ur~Astronomie, K\"onigstuhl~17, 69117~Heidelberg, Germany}

\begin{document}   %
\maketitle

\renewcommand{\thefootnote}{\fnsymbol{footnote}}
\footnotetext[1]{e-mail: {\tt pschuller@cfa.harvard.edu}}
\footnotetext[2]{e-mail: {\tt martin.vannier@unice.fr}}
\renewcommand{\thefootnote}{\arabic{footnote}}

\begin{abstract}
Current detection methods of planetary companions 
do not allow retrieving their spectral information properties. 
The method of Differential Interferometry has the potential 
to complement such information 
by comparing interferometric observables in various spectral channels. 
We outline the basic aspects of this method and how it can soon be realised 
by the Mid-infrared Interferometric instrument (MIDI) 
at the Very Large Telescope Interferometer (VLTI) observatory. 
A set of possible candidates for such direct observation has been selected 
among currently identified planetary companions. 
Differential Interferometry with MIDI would be complementary to other ground-based programs. 
The method may also be an alternative for future space-based planetary spectroscopy. 
\end{abstract}

\section{Introduction}
\label{sect.itro}
For several years, the number of discovered companions 
has been steadily increased, 
mostly by studying their dynamical effects on the host star. 
To date, the majority of the detections use indirect methods 
which do not allow the spectral characterisation of the companions. 
In this contribution, we outline the method of Differential Interferometry, 
applied to gain spectral information on the companions 
and employing interferometric instruments offering moderate spectral resolution. 
One such instrument is the Mid-infrared Interferometric instrument (MIDI), 
which is being commissioned 
	\citep{cit.przygodda2003:dartpf} 
at the Very Large Telescope Interferometer (VLTI) observatory. 
Together with other instruments (like, e.g., AMBER in the near-infrared), 
MIDI could reveal spectra of stellar companions like planets 
already in the medium-term range. 
This would complement current detection methods 
and therefore help prepare future, more versatile projects for observing extra-solar companions. 
Present developments of DI 
	\citep{cit.vannier2003:phd, cit.vannier2003a}, 
if confirmed by soon-to-come observations, 
may indeed push for a space-based use of the differential mode 
in combination with, or as an alternative of, 
the nulling technique foreseen for DARWIN.

\section{Instrumentation}
\label{sect.instruments}

\subsection{The Observatory}
\label{sect.observatory}
\begin{figure*}
	\centering
	\includegraphics[height=0.25\textheight]{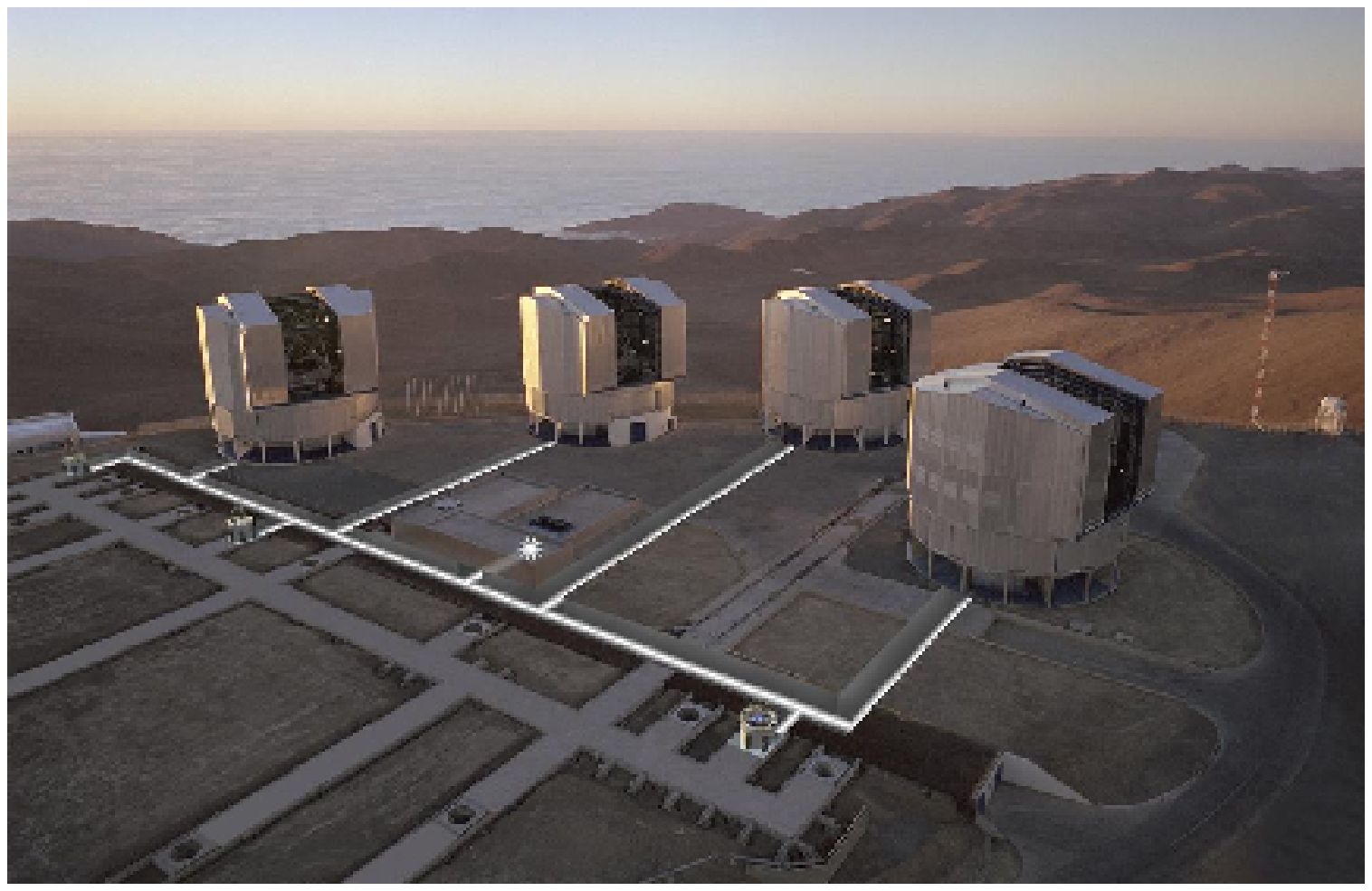}%
		\hfill 
	\includegraphics[height=0.25\textheight]{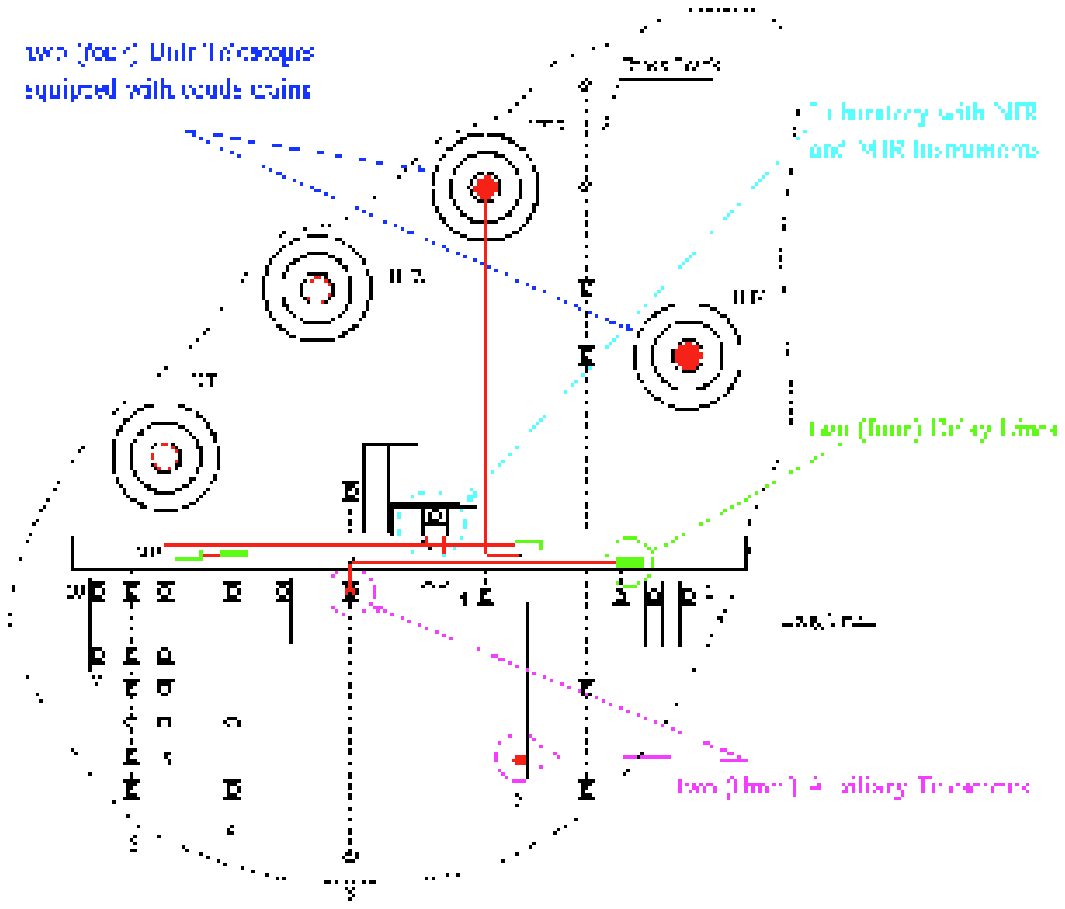}
	\caption{\label{fig.vlti} 
	{Aerial view and ground plan of VLTI.} 
	The aerial view shows the UTs already in place and representations of the ATs which will be installed in 2003. White lines indicate the optical path from the telescopes to the Interferometric Laboratory where the beams are combined. North is up in the layout drawing. 
	(Images: \citep{cit.eso-outreach})   %
		}
	\end{figure*}

The European Southern Observatory (ESO) is currently setting up 
	the VLTI %
on Mt.~Paranal/Chile 
	\citep{cit.vlti2003SPIE, cit.schoeller2003:dartpf}. 
It provides four fixed Unit Telescopes (UTs) with 8.2-m primary mirrors 
and eventually four movable 1.8-m Auxiliary Telescopes (ATs)
which can be relocated to 30 different stations, 
as indicated in Fig.~\ref{fig.vlti}. 
The light from the telescopes is directed through delay line tunnels to a central laboratory. 
Interferometric instruments working in the near- and mid-infrared wavelength regions combine the light of two (or more) telescopes coherently. 
Information on the angular size of an astronomical object is contained in the fringe contrast (visibility) of the interferogram, 
whereas asymmetries are related to phase shifts of the interferogram. 

The spatial resolution $\phi_{res}$ of interferometers is related to 
the separation $B$ of the single telescopes (baseline) 
by $\phi_{res} \propto \lambda / B$. 
At VLTI, telescope separations of the UTs range from $47~\mathrm{m}$ to $130~\mathrm{m}$
and from $8~\mathrm{m}$ to $202~\mathrm{m}$ for the ATs. 
Measuring at a mid-infrared wavelength $\lambda=10~\mathrm{\mu m}$, 
the highest resolution with the UTs is therefore $16~\mathrm{mas}$ and $10~\mathrm{mas}$ for the ATs. 
For comparison, the red giant star $\alpha$\,Ori (Betelgeuse, $60~\mathrm{pc}$ away from Sun) 
shows an apparent diameter of around $55~\mathrm{mas}$ in the mid-infrared 
	\citep{cit.weiner2000}. 

\subsection{The MIDI Instrument}
\label{sect.midi}
\begin{figure*}
	\centering
	\includegraphics[width=0.95\textwidth]{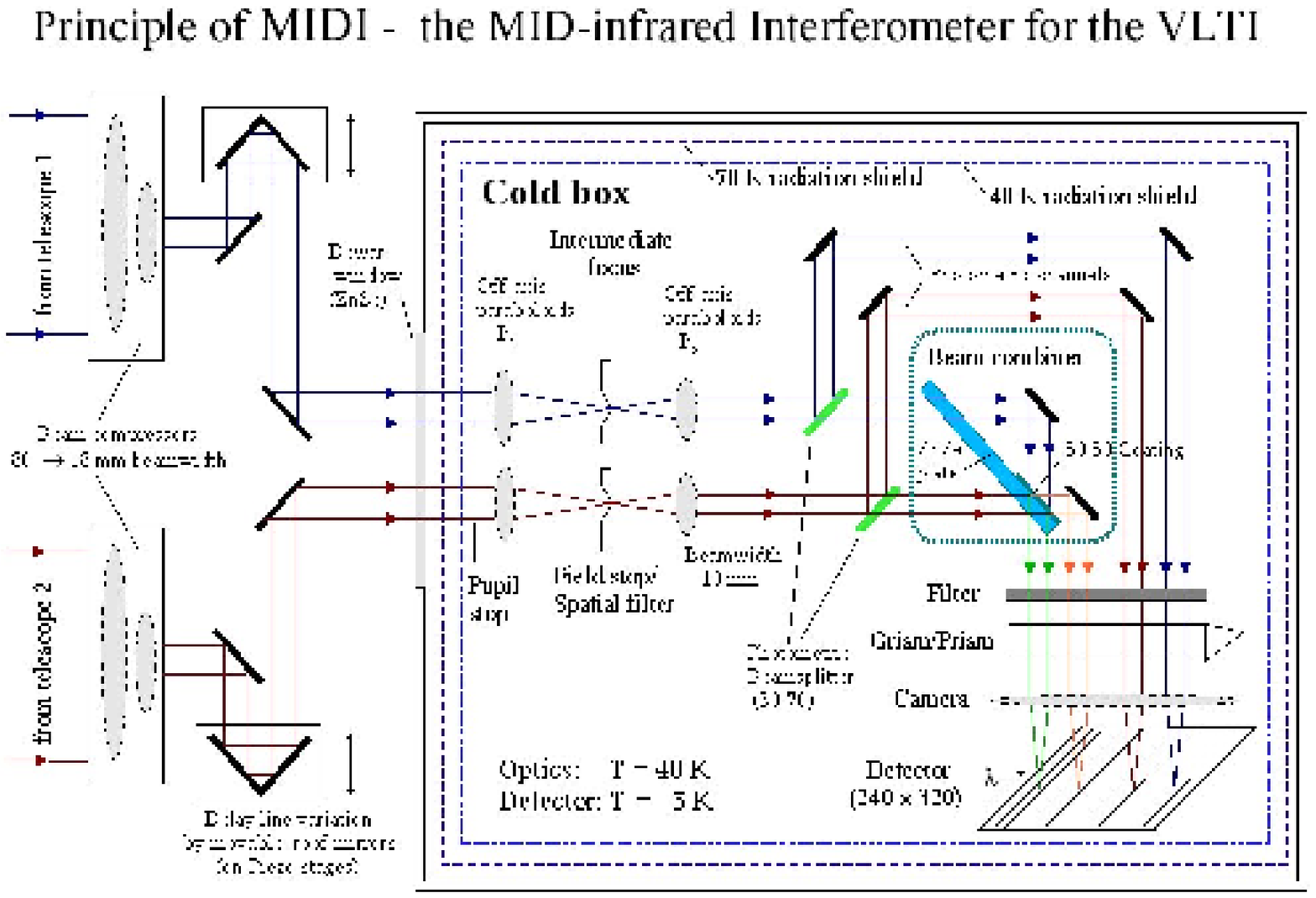}
	\caption{\label{fig.midi} 
	{Scheme of MIDI.} 
	For further explanation of the instrument, 
	refer to \cite{cit.przygodda2003:dartpf} and \cite{cit.midi2003SPIE}. 
		}
	\end{figure*}

MIDI %
	\citep{cit.midi2003SPIE}
is one of the interferometric instruments at the VLTI 
and currently in its commissioning phase 
	\citep{cit.przygodda2003:dartpf}. 
It covers the N band ($8...13~\mathrm{\mu m}$) 
and works as a co-axial pupil plane interferometer with two incoming beams. 
Different spectroscopic and photometric modes are available (see Fig.~\ref{fig.midi}). 
In particular, MIDI provides a grism with a spectral resolution $R_g=(\lambda/\Delta\lambda)=260$ and a prism with $R_p=30$. 
Goals of limiting N magnitudes (for $10\sigma$ detections) are: \\[-4ex] 
	\begin{center}
	\small
	\begin{tabular}{c||c|c}
	\hline\hline
		  &  \multicolumn{2}{c}{Fringe Tracking}  \\
	\cline{2-3}
	Telescopes  &  external  &  self  \\
	\hline\hline 
	UTs  &  9~mag (10~mJy)  &  4~mag (1~Jy)  \\ 
	ATs  &  5.8~mag (200~mJy)  &  0.8~mag (20~Jy)  \\
	\hline
	\end{tabular} .
	\end{center}

Fringe tracking compensates for movements of the interference pattern due to atmospheric disturbances. 
At VLTI, fringe tracking will eventually be performed by FINITO or, respectively, the equivalent sub-unit of the PRIMA instrument 
	\citep[and references therein]{cit.vlti2003SPIE}. 
With the interference pattern stabilised in this way, 
two-beam interferometers like MIDI are able to measure both the modulus $V(B/\lambda)$ 
and the phase $\Phi(B/\lambda)$ (fringe contrast and fringe phase) 
of the complex degree of coherence of the source.

\section{Detection of Sub-stellar Objects}
\label{sect.detect}
In this section we outline a proposal 
prepared for guaranteed time observations with MIDI. 
The project is still under definition, 
and its feasibility will be evaluated on the basis of instrument performance 
as commissioning of MIDI will proceed. 

\subsection{Principle Idea of Differential Interferometry}
\label{sect.idea}
Spectral dispersion of the interferometric signal in MIDI 
brings the possibility of direct extra-solar planet detection 
using Differential Interferometry (DI) 
	\citep{cit.di2000SPIE}. 
For a given orbital configuration at a time $t$ of a binary system with a period $P$, 
a dispersed interferogram contains the displacement of the photo-center 
dependent on wavelength. 
Thus, it yields both the system's angular separation 
and the dependence of the planet/star luminosity ratio on wavelength. 
DI is based on comparing measurements 
of visibility $V(t/P,B/\lambda)$ and/or phase $\Phi(t/P,B/\lambda)$ 
at different wavelengths to a reference wavelength, 
and so canceling all non-chromatic systematic instrumental errors. 
Depending on the resolution chosen for the dispersion, 
the measured spectral features should help constrain 
the parameters of radiative equilibrium of the planetary atmosphere, 
in particular its temperature and chemical composition. 

\subsection{Parameters}
\label{sect.params}
\begin{figure*}
	\centering
	\includegraphics[width=0.5\piccol]{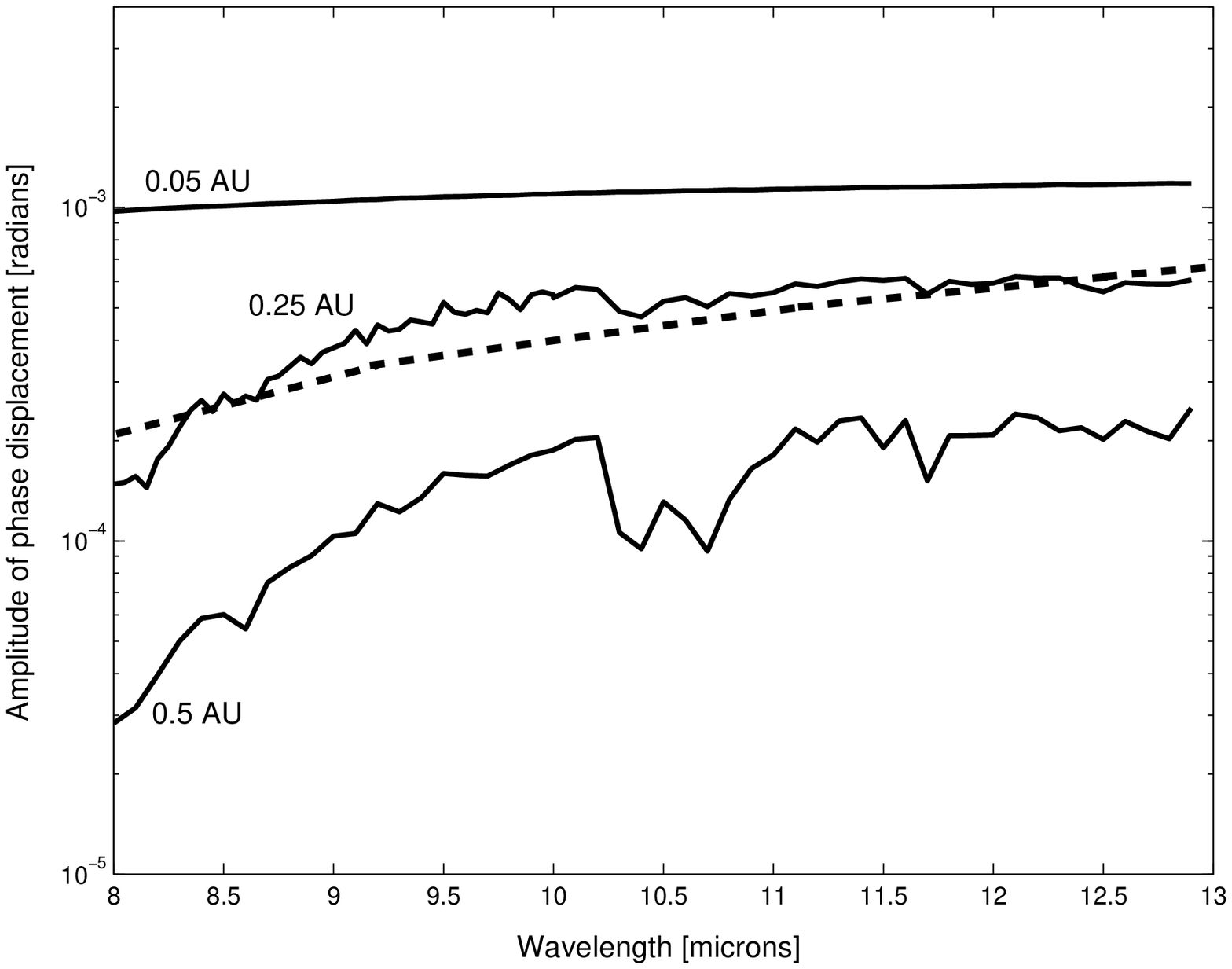}
	\includegraphics[width=0.5\piccol]{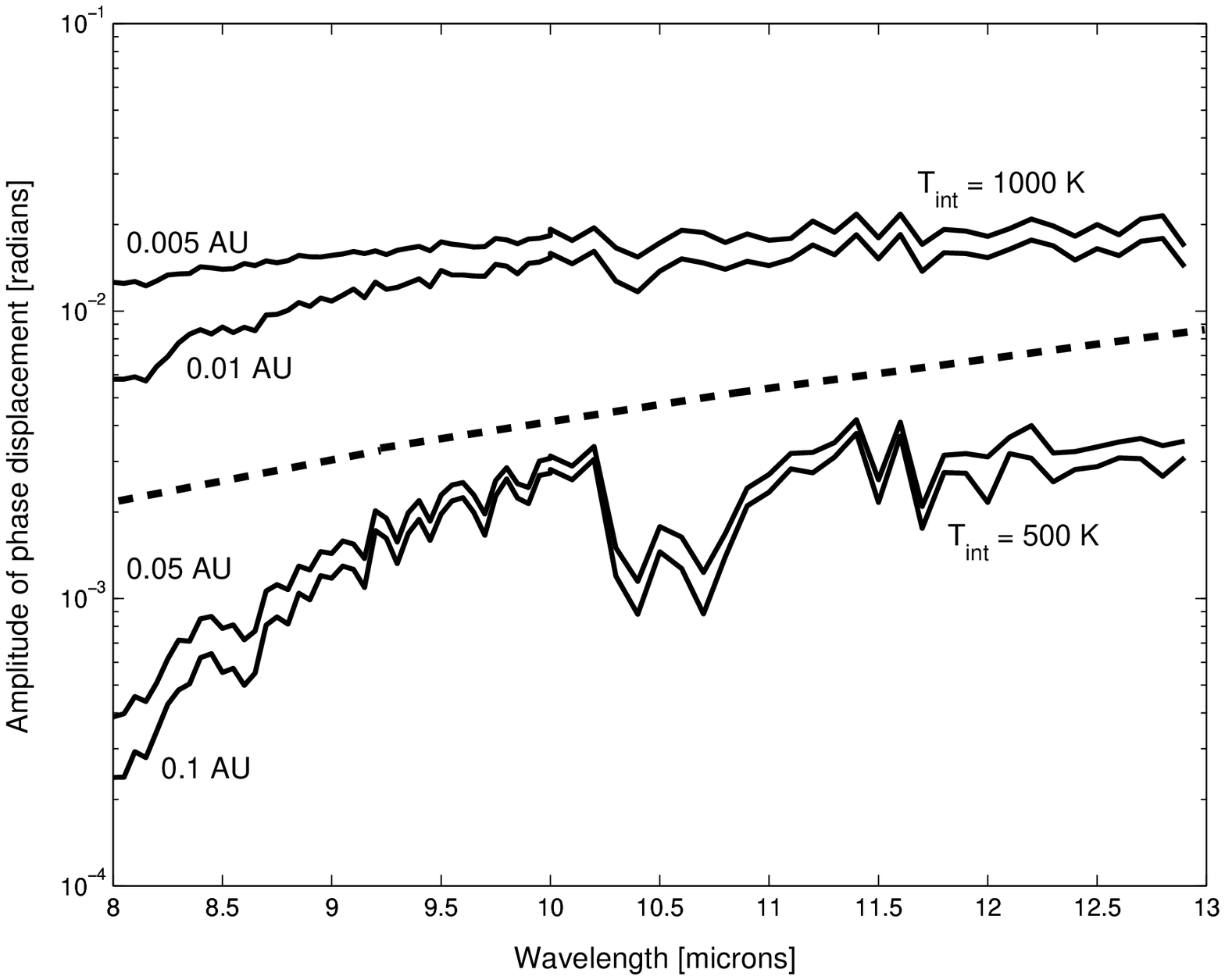}
	\caption{\label{fig.phasevar} 
	{Expected amplitude of phase shifts with wavelength of a star + planet system located at $10~\mathrm{pc}$, for various orbital distances.} 
Shifts are with respect to a star without any companion. 
The synthetic spectrum of the planetary companion assumes a 1-Jupiter mass body and a non-dusty atmosphere (AMES-Cond simulations 
	\citep{cit.barman2001}). 
Noise levels (dashed lines) are for a 3-hour observation on 2 UTs with MIDI, a spectral resolution of 30, and the use of photometric channels. %
{Left}: G2 star; {Right}: M5 star.
		}
	\end{figure*}

The theoretical limitation for DI observations is set by the fundamental noise level. 
For MIDI, it is mainly carried by the thermal background noise 
(the photon noise from the source and the read-out noise being minor contributions at $10~\mathrm{\mu m}$). 
The error of a %
	visibility or phase measurement 
is a function of collecting area, spectral resolution, observing time, and brightness of the object. 
For example, observing a G2 star at $10~\mathrm{pc}$ ($N=2.7$) 
using two UTs during 3 hours and with a spectral resolution of 30 (see Sect.~\ref{sect.midi}) 
yields a noise error equivalent to a phase shift 
of $2\ldots 7\cdot10^{-4}~\mathrm{rad}$ (rms), depending on wavelength, 
whereas an M5 star at the same distance ($N=6.4$) 
has an error of $2\ldots 9\cdot10^{-3}~\mathrm{rad}$ (rms). 
In Fig.~\ref{fig.phasevar}, these errors are indicated 
by a bold dashed line in the respective plot. 

The detection potential for exo-planets appears 
when comparing their induced variations of the visibility or phase 
to the noise levels. 
Fig.~\ref{fig.phasevar} shows, as a function of wavelength, 
the simulated shifts of the phase $\Phi$ 
due to the presence of a Jupiter-size planet, 
as compared to the case without any companion. 
These displacements are plotted 
for two host stars of different spectral type 
and, in each case, for several semi-major axes of the planet orbit. 
Observations were considered at the maximum planet-star angular separation 
and with the direction of the interferometer baseline being parallel to the separation vector. %
We used the synthetic spectra from 
	\citep{cit.barman2001}. 
The applied albedo of the planet is strongly dependent on the wavelength. 
The planet's rotation plays a minor role in the obtained results, 
as does the intrinsic temperature for highly irradiated planets. 
The planet/star luminosity ratio, and so the interferometric signature, 
should grow for closer orbits 
and higher intrinsic temperatures of a companion 
(the latter depending on its mass and age).
Generally, the method of DI applied with MIDI on VLTI/UTs 
(preferably with the longest possible baseline) 
potentially resolves, and allows to study, systems like 51~Peg, 
i.e., a Sun-like star at $10~\mathrm{pc}$ 
and a Jupiter-like planet orbiting with a semi-major axis smaller than $0.07~\mathrm{AU}$, 
and possibly planetary systems around M-type stars with extremely small orbital radii. 
Notably, brown dwarfs are not only larger, 
but also have a much higher thermal emission, 
independent of their distance to the parent star. 
They are expected to be good candidates for direct detection with DI. 

An additional practical limitation to the potential of detection 
might arise from the chromatic Optical Path Difference (OPD) between the beams, 
induced by dispersion in the transmissive media passed by the beams. 
Main sources for the chromatic OPD bias are 
the air in the delay line, the dioptrics of VLTI, 
and some possible deformation of the dispersive elements or the detector of MIDI. 
Since the latter, static effects can in principle be calibrated using a reference object, 
the relevant bias is, in fact, the evolution of the chromatic OPD with time 
relative to the calibration period. 
Reference~\citep{cit.meisner2003SPIE} indicates 
a strong effect of dispersion in the mid-infrared 
and proposes solutions for at least partial correction. 
Estimates have been made on the amplitude and on the time constants of each of the different contributions. 
We anticipate that, at $10~\mathrm{\mu m}$, the corrected OPD effects can be lowered below the noise levels. 
This must imply, in particular, monitoring the dispersion 
due to the variable humidity in the atmosphere. 

\subsection{Target Selection}
\label{sect.targets}
Candidates for the described method come mainly from radial velocity detections. 
The measurements provide the orbit's semi-major axis, its period and orbital phase, 
and a lower estimate $M\cdot\sin i$ of the companion's mass, where $i$ is the inclination angle of the orbital plane. 
The behaviour of a planet's size versus its mass is highlighted, 
for example, in \cite{cit.burrows2000}. 

Targets suitable for MIDI must have, first of all, a sufficient overall brightness. 
They should be as large and as warm as possible 
in order to have a favourable flux ratio between star and companion. 
Smaller angular separations, increasing thermal emission, 
need to be balanced with 
wider separations which are better resolved by the interferometer.

\section{Conclusions}
\label{sect.concl}
Differential Interferometry appears to be a promising method 
to study atmospheric characteristics of 51-Peg-like planets, 
i.e., hot and massive extra-solar planets 
located closer than $0.07~\mathrm{AU}$ from their star. 
Currently, three sub-stellar companions detected by indirect methods 
provide conceivable candidates for direct detection around $10~\mathrm{\mu m}$ 
(GL~86, HD~112758, HD~217580). %
Though challenging in some instrumental aspects, 
and therefore feasibility needs to be evaluated along instrument commissioning, 
this method could be soon performed successfully with MIDI. 
If successful, it will be complementary to a similar DI observation program 
proposed with the near-infrared instrument AMBER on VLTI, 
which may also be performed in the near future. 

In a broader perspective, 
the differential phase method may be foreseen for space-based applications. 
In the case of direct detection and spectroscopy of extra-solar planets, 
the space segment offers an access to the IR with considerable advantages: 
optimal planet/star flux ratio with low background noise and no chromatic dispersion due to the atmosphere. 
It may then bring a valuable alternative 
to nulling interferometry as a detection technique. 
There would be then no challenging need for stellar extinction, 
but it would require severe control of the chromatic OPD 
along the instrumental chain. 
Further studies would be worthwhile estimating the practical feasibility 
and performance of this method as a space technique.

   %

\end{document}